# GNSS/GPS Spoofing and Jamming Identification Using Machine Learning and Deep Learning

Ali Ghanbarzadeh, Muhammad Soleimani and Hossein Soleimani*

School of Electrical Engineering, Iran University of Science and Technology, Tehran, Iran

*Corresponding Author (hsoleimani@iust.ac.ir)

**Abstract**: The increasing reliance on Global Navigation Satellite Systems (GNSS), particularly the Global Positioning System (GPS), underscores the urgent need to safeguard these technologies against malicious threats such as spoofing and jamming. As the backbone for positioning, navigation, and timing (PNT) across various applications—including transportation, telecommunications, and emergency services—GNSS is vulnerable to deliberate interference that poses significant risks. Spoofing attacks, which involve transmitting counterfeit GNSS signals to mislead receivers into calculating incorrect positions, can result in serious consequences, from navigational errors in civilian aviation to security breaches in military operations. Furthermore, the lack of inherent security measures within GNSS systems makes them attractive targets for adversaries. While GNSS/GPS jamming and spoofing systems consist of numerous components, the ability to distinguish authentic signals from malicious ones is essential for maintaining system integrity. Recent advancements in machine learning and deep learning provide promising avenues for enhancing detection and mitigation strategies against these threats. This paper addresses both spoofing and jamming by tackling real-world challenges through machine learning, deep learning, and computer vision techniques. Through extensive experiments on two real-world datasets related to spoofing and jamming detection using advanced algorithms, we achieved state-of-the-art results. In the GNSS/GPS jamming detection task, we attained approximately 99% accuracy, improving performance by around 5% compared to previous studies. Additionally, we addressed a challenging tasks related to spoofing detection, yielding results that underscore the potential of machine learning and deep learning in this domain.

All the codes and experiments related to our research can be found on our GitHub page at the following link: https://github.com/alicmu2024/GNSS-Jamming-Detection-and-Classification-using-Machine-Learning-Deep-Learning-and-Computer-Vision/tree/main.

## 1. INTRODUCTION

Global Navigation Satellite Systems (GNSS) are essential to modern society, providing critical positioning, navigation, and timing (PNT) services that drive economic growth across various sectors. Key constellations, including GPS, GLONASS, Galileo, and BeiDou, enable precise navigation for vehicles, aircraft, and maritime vessels, ensuring safe and efficient travel [1]. In agriculture, GNSS facilitates precision farming by enabling field mapping and automated machinery operation, thereby increasing productivity. Furthermore, GNSS plays a crucial role in telecommunications and financial services by synchronizing networks and transactions. As this technology evolves, its applications are expanding into new domains such as autonomous vehicles and the Internet of Things (IoT), where real-time location awareness is vital for enhancing the functionality of smart technologies [1, 2]. Despite their importance, GNSS systems face significant threats from malicious activities such as jamming and spoofing [3]. Jamming involves the deliberate transmission of signals that interfere with legitimate GNSS signals, rendering them unusable. This disruption can compromise critical services reliant on accurate positioning and timing, leading to severe safety risks across various sectors, including transportation and emergency services [4, 5]. Conversely, spoofing represents a more sophisticated form of attack in which counterfeit GNSS signals deceive receivers into calculating incorrect positions. The absence of built-in security features in GNSS systems exacerbates their vulnerability, leaving many applications exposed to potential risks. To effectively detect defects in GNSS signals, it is crucial to differentiate between authentic and non-authentic signals to maintain the integrity of applications susceptible to spoofing and jamming attacks [6]. The detection process begins with signal acquisition, where the receiver identifies available satellites and continues with signal tracking to ensure alignment. The core detection mechanism involves analyzing signal characteristics such as strength and correlation patterns to identify discrepancies indicative of spoofing or jamming; for instance, spoofing attacks often produce multiple correlation peaks compared to the single peak typical of genuine signals [7]. Traditional methods for this analysis have limitations, relying on predefined thresholds that may not be effective in all scenarios and are prone to false alarms [4, 5]. In contrast, machine learning (ML) and artificial intelligence (AI) methods enhance detection by analyzing patterns in large datasets of known signals, recognizing subtle differences that traditional methods might miss. These advanced techniques can adapt to complex scenarios, improving detection accuracy over time. However, recent advancements in detecting jamming and spoofing attacks in GNSS using ML and deep learning techniques reveal notable shortcomings. Most studies focus exclusively on either spoofing or jamming, neglecting the simultaneous occurrence of both threats, which limits detection robustness[3-5, 8]. Additionally, many methodologies are tested under controlled conditions or specific threat models, failing to generalize to real-world scenarios. Reliance on predefined thresholds leads to high rates of false positives and negatives since traditional methods do not adapt to evolving attack strategies[9-11]. While ML and deep learning have the potential for adaptation, many implementations remain simplistic. Furthermore, there is a lack of validation across various GNSS constellations; most research concentrates on specific systems like GPS, limiting the applicability of detection methods [11, 12]. In this research, we investigate the efficacy of a diverse array of machine learning and deep learning algorithms for detecting and identifying GPS spoofing and various forms of jamming. This exploration highlights both the strengths and weaknesses of these algorithms in analyzing complex GNSS data. Our methodologies were rigorously tested across three distinct datasets: two focused specifically on spoofing scenarios while the third concentrated on jamming detection. This comprehensive approach allows for a nuanced understanding of the capabilities and limitations inherent in current detection techniques. To the best of our knowledge, this study represents the first comprehensive investigation that simultaneously addresses both GPS spoofing and jamming. Through extensive experiments, we provide a thorough analysis of the detection capabilities of various machine learning and deep learning algorithms alongside computer vision techniques in these critical areas. This pioneering approach contributes significantly to the existing body of research by offering insights into the interplay between these two types of attacks and their implications for GNSS security.

The remainder of this paper is organized as follows: In Section 2, we review the related works on jamming and spoofing detection in GNSS, highlighting existing methodologies and their limitations. In Section 3, we detail the machine learning and deep learning algorithms employed in our study. Section 4 presents the datasets used for testing, describing their characteristics and relevance to our research objectives. In Section 5, we analyze the results of our experiments, discussing the performance of various algorithms in detecting GPS spoofing and jamming. Finally, we conclude the paper in Section 6, summarizing our findings and suggesting directions for future research.

## 2. RELATED WORKS

### 2.1. Spoofing Detection Using Machine Learning

GPS spoofing is a malicious technique that involves the manipulation of Global Positioning System (GPS) signals to deceive a receiver into calculating an inaccurate position. This is accomplished by broadcasting counterfeit signals that overpower the legitimate signals transmitted by GPS satellites, resulting in the receiver erroneously believing it is located at an alternative position. The potential implications of GPS spoofing are significant, as it can disrupt navigation systems and compromise the integrity of various applications that rely on accurate location data [13].

Numerous studies have been conducted to detect, identify, and mitigate GPS spoofing attacks on unmanned aerial vehicles (UAVs). The authors of [14] proposed a classification technique using artificial neural networks (ANNs), incorporating benchmarks such as signal-to-noise ratio, pseudo-range, and Doppler shift. The authors of [15] employed Linear Regression and Long Short-Term Memory (LSTM) for detection, evaluating their model based on time steps and neurons. Support Vector Machine (SVM) algorithms were explored by the authors of [9], who evaluated their performance based on evaluation windows and time width. A deep learning method called DeepSIM was proposed by the authors of [16], which detects spoofing by comparing historical GPS images to incoming images using image processing techniques. The authors of [17] introduced two dynamic selection approaches based on ten commonly used machine learning models. The authors of [18] compared tree-based models—Extreme Gradient Boosting (XGBoost), Random Forest (RF), Gradient Boosting (GBM), and Light Gradient Boosting (LightGBM)—for detecting GPS spoofing in UAVs, utilizing a benchmark with 13 features. Another study by the authors of [19] combined SVM with K-fold cross-validation for enhanced detection. The authors of [20] proposed a semantic-based detection technique called CONSDET for onboard GPS spoofing detection. Despite these advancements, most research has primarily focused on specific machine learning and deep learning models without adequately addressing the influence of dataset characteristics and model parameters on performance. Challenges such as small or biased datasets can hinder model generalization and effectiveness. Moreover, obtaining high-quality data reflecting GPS spoofing attacks remains difficult; thus, enhancing dataset quality is crucial for improving detection outcomes. Further investigation into these factors is necessary to establish best practices and guide future research directions in GPS spoofing detection techniques.

In recent studies, the performance of various supervised and unsupervised models for detecting GPS spoofing attacks has been rigorously evaluated. The researchers compared several supervised models against unsupervised ones, assessing metrics such as accuracy, detection probability, and processing time. Their findings revealed that classification and regression decision tree models outperformed other algorithms in detecting and classifying GPS spoofing attacks [21].

The paper in [22] provides a comprehensive survey of machine learning (ML) techniques aimed at enhancing Global Navigation Satellite System (GNSS) positioning accuracy. It discusses various ML methods, including supervised, unsupervised, deep learning, and hybrid approaches, and their applications in signal analysis, anomaly detection, multi-sensor integration, and prediction.

### 2.2. Jamming Detection

Traditional methods for detecting GPS jamming primarily encompass techniques such as signal filtering, directional antennas, and advanced algorithms [23]. While these methods for GPS jamming detection have laid the groundwork for anti-jamming technologies, their limitations—particularly in high-jamming environments—underscore the need for more robust solutions, such as those offered by machine learning and deep learning techniques [25]. Recent advancements in machine learning techniques have significantly enhanced the detection and classification of jamming signals in Global Navigation Satellite Systems [9, 25]. These techniques demonstrate considerable potential for improving resilience against radio frequency interference. Machine learning approaches have been effectively employed to detect and classify jamming attacks, particularly against orthogonal frequency division multiplexing systems [25]. Additionally, a deep reinforcement learning model has been developed to address anti-jamming scenarios in both static and dynamic environments [26]. In the realm of deep learning, convolutional neural networks combined with long short-term memory networks have been utilized to effectively detect transient radio frequency, enabling precise identification of interference sources [27]. Furthermore, bidirectional long short-term memory networks have been implemented specifically for signal interference detection [28]. Notably, an attention mechanism integrated with recurrent neural networks has been introduced to enhance throughput prediction for long-term evolution systems. These advancements underscore the growing importance of machine learning and deep learning techniques in improving the robustness of GNSS and wireless communication systems against jamming attacks [ 29].

## 3. PROBLEM STATEMENT

The increasing reliance on Global Navigation Satellite Systems (GNSS), particularly the Global Positioning System (GPS), for various applications has made them vulnerable to spoofing and jamming attacks. Spoofing involves broadcasting false signals that mislead receivers into calculating incorrect positions, while jamming disrupts the legitimate signals, rendering GNSS unusable. These vulnerabilities pose significant risks to safety-critical systems, including transportation, logistics, and emergency services. Despite existing detection and mitigation strategies, many current approaches lack effectiveness in real-world scenarios due to their dependence on specific conditions or the availability of additional hardware. Moreover, traditional methods often fail to adapt to the dynamic nature of jamming and spoofing attacks, particularly as attackers develop more sophisticated techniques. Consequently, there is a pressing need for robust detection mechanisms that can operate effectively in diverse environments. This research endeavors to tackle the pressing challenges associated with GNSS spoofing and jamming attacks by leveraging advanced machine learning (ML) and deep learning (DL) algorithms. By emphasizing automated feature extraction and the inherent pattern recognition capabilities of these techniques, this study aims to establish a comprehensive framework that significantly enhances resilience against such threats. A primary motivation for this research arises from the observation that most existing studies have concentrated exclusively on either spoofing detection or jamming detection within GNSS contexts, often employing a limited selection of ML and DL algorithms. In contrast, our research innovatively addresses both tasks concurrently, utilizing a diverse array of sophisticated algorithms. This approach not only demonstrates the efficacy of these methods in managing the complexities associated with GNSS vulnerabilities but also underscores their potential to revolutionize detection strategies in this domain. Moreover, we have made all our code and experimental data publicly accessible, thereby contributing a valuable benchmark for future research in this field. This commitment to transparency not only facilitates reproducibility but also fosters further exploration of effective solutions to GNSS security challenges. Additionally, our work incorporates advanced training techniques for both ML and DL algorithms across each task, yielding state-of-the-art results that set a new standard for performance in the field. Through these contributions, this research not only fills existing gaps in the literature but also paves the way for future advancements in GNSS security, ultimately enhancing the integrity and reliability of critical applications reliant on these systems.

## 4. DATASETS

In this paper, we utilize two distinct types of datasets: one for detecting GPS spoofing in aerial vehicles and another for jamming detection. This section will provide a detailed discussion of these datasets, outlining their characteristics and relevance to our research objectives.

### 4.1. A Dataset for GPS Spoofing Detection on Unmanned Aerial System

This dataset comprises data extracted from authentic GPS signals collected using an eight-channel GPS receiver mounted on a vehicle [30]. The vehicle was operated at speeds ranging from 0 mph to 60 mph to simulate the flight dynamics of unmanned aerial systems (UAS). Additionally, data was collected from three static positions atop various buildings at different altitudes to emulate the hovering behavior of UAS. The dataset for GPS spoofing detection includes 13 distinct features, each providing critical information for identifying and classifying GPS signals. These features are as follows:

1. **PRN (Pseudo-Random Noise)**: This feature represents the unique identifier for each satellite in the GPS constellation, allowing differentiation between signals from multiple satellites.
2. **DO (Doppler Offset)**: This parameter measures the frequency shift of the received signal due to relative motion between the satellite and the receiver, which can indicate potential spoofing.
3. **PD (Pseudo-Range)**: This is the calculated distance between the GPS receiver and the satellite, derived from the time delay of the received signal, essential for accurate positioning.
4. **RX (Receiver Time)**: This feature captures the time at which the signal is received, providing a temporal context that is crucial for synchronization and positioning calculations.
5. **TOW (Time of Week)**: This indicates the specific time within the week when the signal was transmitted, which helps in assessing the validity of the received data.
6. **CP (Carrier Phase)**: This feature measures the phase of the received carrier wave, providing additional

information about signal integrity and aiding in precise positioning.
7. **EC (Elevation Angle)**: The angle of the satellite above the horizon relative to the receiver, which can affect signal strength and quality.
8. **LC (Lock Condition)**: This parameter indicates whether the receiver has successfully locked onto a satellite signal, reflecting its ability to maintain a stable connection.
9. **PC (Packet Count)**: This feature counts the number of packets received over a specific period, providing insights into signal continuity and potential disruptions.
10. **PIP (Position Information Packet)**: This feature contains critical data about the position information being transmitted by the satellite, essential for accurate navigation.
11. **PQP (Position Quality Parameter)**: This parameter assesses the quality of position data based on various metrics, helping to identify unreliable signals.
12. **TCD (Time Correction Data)**: This feature provides adjustments to account for timing discrepancies between satellites and receivers, crucial for maintaining accuracy.
13. **CN0 (Carrier-to-Noise Ratio)**: This metric measures the strength of the received signal relative to background noise, indicating signal quality and potential interference.

The target variable in the dataset for GPS spoofing detection is designed to classify each signal instance as either legitimate or spoofed, reflecting the effectiveness of the detection algorithms. In addition to authentic GPS signals, three distinct types of GPS spoofing attacks were simulated: simplistic, intermediate, and sophisticated.

- **Simplistic attacks** involve the transmission of fake GPS signals that are unsynchronized with authentic signals, often resulting in higher Doppler shift measurements and a significant signal-to-noise ratio due to their elevated power levels.
- **Intermediate attacks** are characterized by the attacker having knowledge of the UAV's position, allowing for the generation of GPS signals that maintain Doppler shift measurements and pseudo-range values within normal ranges.
- **Sophisticated attacks** represent the most complex threat, where the spoofer gains control over multiple synchronized antennas, manipulating several channels simultaneously. This type of attack is particularly challenging to detect due to the effects of multipath signals and the dynamic nature of satellite motion. [A DATASET for GPS Spoofing Detection on Autonomous Vehicles].

## 4.2. Raw IQ dataset for GNSS/GPS Jamming Signal Classification

**Overview of the jamming detection dataset:** The dataset utilized for GNSS jamming signal classification is a comprehensive image dataset derived from Raw In-phase and Quadrature (IQ) data, designed to analyze the impact of various jamming signals on GPS receivers [31]. It comprises six distinct classes, each representing a different type of jamming signal, and is formatted as spectrogram images that facilitate effective analysis and classification using machine learning and deep learning techniques.

**Conversions of the jamming signals to images:** To effectively analyze jamming signals in GNSS applications, raw In-phase and Quadrature (IQ) data is transformed into visual representations, specifically spectrogram images. This conversion process begins with the collection of IQ samples, which capture the amplitude and phase information of the received signals over time. The next step involves applying a time-frequency transformation, such as the Short-Time Fourier Transform (STFT), to convert the 1D time-domain signal into a 2D time-frequency domain representation. This transformation produces a spectrogram that visually illustrates how the frequency content of the signal varies over time, allowing for easier identification and classification of different jamming types. The resulting images serve as input for machine learning and deep learning models, enabling the classification of jamming signals based on their unique spectral features.

**Description of Jamming Signal Classes:** The dataset for GNSS jamming signal classification comprises six distinct classes, each representing a different type of jamming signal:
1. **SingleAM**: This class represents jamming signals that utilize amplitude modulation (AM) techniques. These signals can effectively overpower legitimate GPS signals by varying their amplitude, making them particularly disruptive in environments where signal strength is critical.
2. **SingleChirp**: This class includes chirp jamming signals that sweep across frequencies over time. Chirp jammers are known for their high effectiveness due to their ability to cover a wide frequency range, which can confuse GPS receivers by introducing rapidly changing frequencies.
3. **NB (Narrowband)**: Narrowband jamming signals are characterized by their limited bandwidth and are designed to interfere with specific frequency channels. This class targets particular frequencies used by GNSS signals, potentially causing significant disruptions in navigation accuracy.
4. **NoJam**: This class serves as the baseline category, representing scenarios where no jamming is present. It includes authentic GNSS signals without any interference, providing a reference point for distinguishing between legitimate and jamming signals.
5. **SingleFM**: This class encompasses frequency modulation (FM) jamming signals that modulate the frequency of the carrier wave to disrupt legitimate GPS signals. FM jammers can create complex interference patterns that challenge traditional detection methods.

6. **DME (Distance Measuring Equipment)**: This class includes signals from Distance Measuring Equipment, which can sometimes overlap with GNSS frequencies. While not strictly a jamming signal, it can interfere with GNSS operations if not properly accounted for in signal processing.

## 5. EXPERIMENTS

In the experiments conducted for this research, proper preprocessing steps were implemented for both the GPS spoofing and jamming detection datasets to ensure data quality and enhance model performance. Ultimately, we reported various metrics for both tasks, including accuracy, precision, recall, F1-score, and area under the receiver operating characteristic curve (AUC-ROC).

### 5.1. GPS Spoofing Detection

The dataset utilized for GPS spoofing detection comprises a substantial collection of 510,530 non-null samples, each characterized by 13 numerical features. These samples are organized into four distinct classes, as previously discussed in the earlier sections.

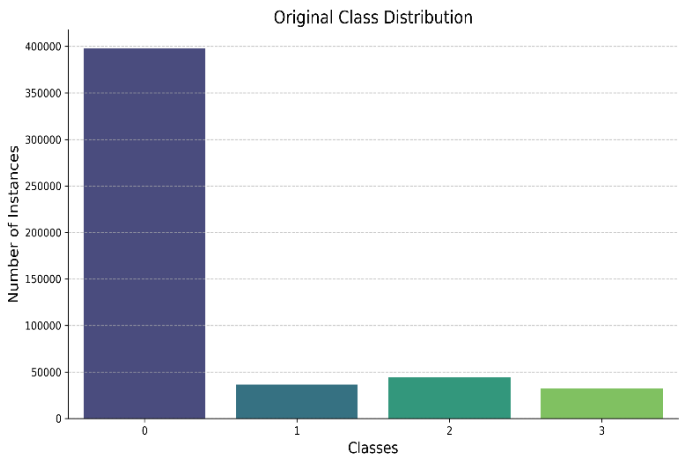

**Figure 1:** Class distribution in the GPS spoofing detection dataset, illustrating significant class imbalance.

Through exploratory data analysis, we discovered that the dataset is significantly imbalanced, with certain classes having a disproportionately higher number of samples compared to others. To illustrate this imbalance, we plotted the distribution of samples for each class, as shown in Figure 1. To address the class imbalance issue, we implemented three distinct methods: random oversampling, random undersampling, and the Synthetic Minority Over-sampling Technique (SMOTE).

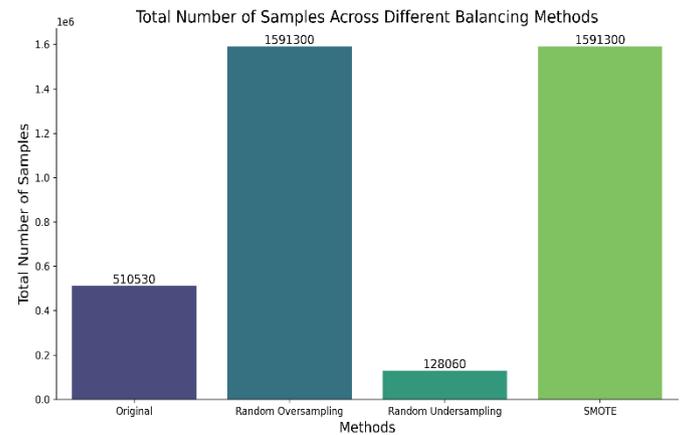

**Figure 2:** Comparison of total sample sizes for the original dataset and the datasets after implementing various balancing methods.

Random oversampling involves the duplication of instances from the minority classes to achieve a more balanced dataset. In contrast, random undersampling reduces the number of samples from the majority classes to mitigate their dominance. SMOTE, however, takes a more advanced approach by generating synthetic samples for the minority classes through interpolation between existing instances.

In Figure 2, we present a comparison of the total number of samples in the original dataset alongside the samples generated through our implemented methods. As illustrated in the figure, techniques such as random oversampling and SMOTE significantly increase the total number of samples, resulting in a substantially larger dataset. Given this substantial growth, we ultimately opted to utilize the dataset after applying random undersampling as our final dataset for model training. This decision was made to maintain a manageable dataset size while ensuring that the class distribution was more balanced, thereby enhancing the effectiveness of our classification algorithms. Following the preprocessing steps, we divided the dataset into two parts: 70% of the data was allocated for training, while the remaining 30% was reserved for testing.

We employed a diverse array of machine learning algorithms to tackle the GPS spoofing detection task, including Logistic Regression, K-Nearest Neighbors (KNN), Gaussian Naïve Bayes, Support Vector Machines (SVM), Decision Tree Classifier, Random Forest Classifier, and Gradient Boosting Machine (GBM) utilizing XGBoost. Each algorithm was systematically evaluated to compare its performance in accurately detecting spoofing attacks, allowing us to identify the most effective models for this specific application. This comprehensive evaluation provides valuable insights into the strengths and weaknesses of each algorithm in addressing the complexities of the dataset.To enhance the training process, we utilized a stratified shuffle split technique in conjunction with grid search cross-validation. This combined approach ensures that each training and testing split maintains the original class distribution, thereby preserving the representativeness of the dataset. Additionally, grid search cross-validation enables us to systematically explore a range of hyperparameter combinations for each algorithm, optimizing their performance by identifying the most suitable parameter settings. This methodical strategy contributes to the robustness and reliability of our model evaluations in detecting GPS spoofing attacks. We evaluated each model on our test sample and found that the Gradient Boosting Machine (GBM) utilizing XGBoost outperforms all other models, achieving an impressive accuracy of approximately 94.44%. In Figure 3, we present the confusion matrix for our best-performing model, the Gradient Boosting Machine (GBM) utilizing XGBoost.

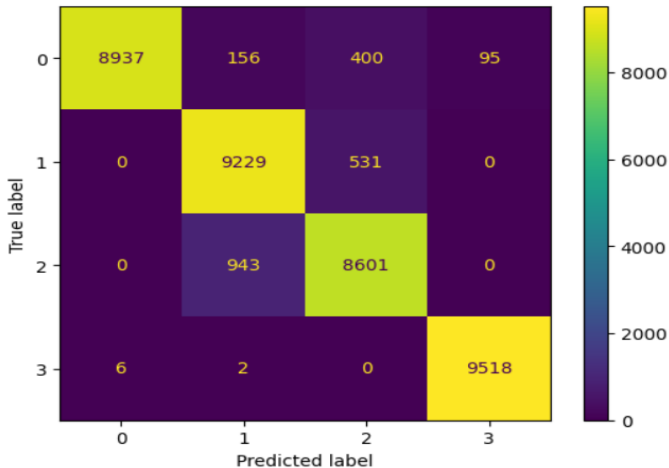

**Figure 3:** Confusion matrix for the best-performing model, illustrating classification performance across all classes.

The second and third best-performing models were the Decision Tree Classifier and the Random Forest Classifier, with accuracies of 94.33% and 94.22%, respectively. These results highlight the effectiveness of GBM with XGBoost in accurately detecting GPS spoofing attacks, while also demonstrating that traditional models like Decision Trees and Random Forests can perform competitively in this context. Another metric that we utilized for comparing models was the Receiver Operating Characteristic Area Under the Curve (ROC AUC), which provides a comprehensive evaluation of a classifier's performance across various thresholds. The ROC curve plots the True Positive Rate (TPR) against the False Positive Rate (FPR), illustrating the trade-offs between sensitivity and specificity. A higher ROC AUC score indicates better model performance, with a perfect model achieving an AUC of 1, while a random model would have an AUC of 0.5. We visualized the ROC AUC for our best model in Figure 4. From the mentioned figure, we observe that the ROC AUC scores for classes 0, 1, and 2 are all approximately 0.99, while class 3 achieves a perfect score of 1.0. These results indicate that the model demonstrates exceptional performance in distinguishing between legitimate signals and spoofing attacks across these classes.

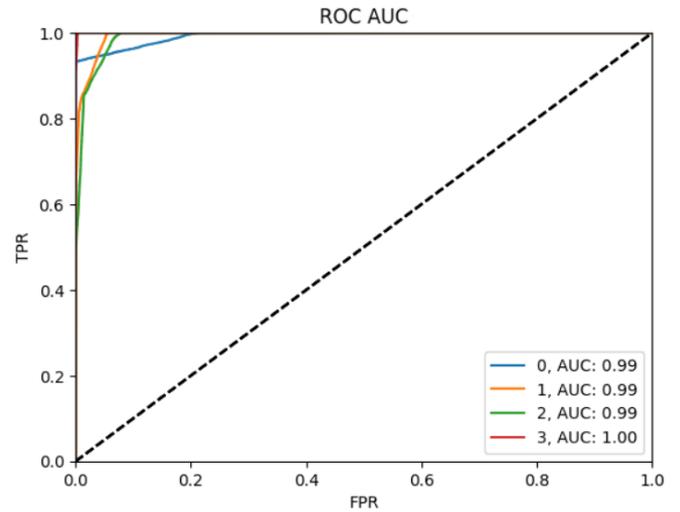

**Figure 4:** ROC AUC curves for each class, showcasing the model's performance in distinguishing between legitimate signals and spoofing attacks.

For each model, we reported all relevant metrics; however, to maintain brevity in our paper, we have included only the results of our best-performing model. The general comparison of the accuracy of each model on both validation and test samples is illustrated in Figure 5. As depicted in the figure, there is no indication of overfitting among the models, as their performance remains consistent across both datasets. This consistency suggests that the models are generalizing well to unseen data, thereby reinforcing the reliability of our findings.

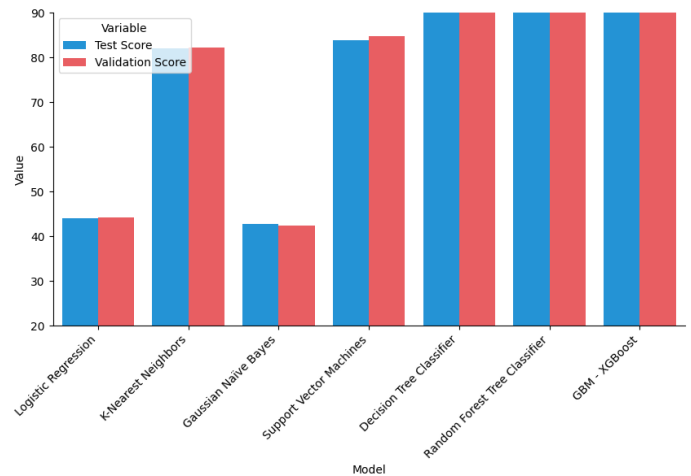

**Figure 5:** Comparison of accuracy for each model on validation and test samples, indicating no signs of overfitting.

For detailed information regarding the performance of all models and their respective metrics, we refer readers to the project's GitHub page. This resource provides comprehensive insights into the methodologies employed, the datasets used, and the complete results of our evaluations, allowing for further exploration and understanding of the findings presented in this study.

## 5.2 Jamming Signal Classification

The dataset utilized for jamming detection is the Raw IQ dataset, which comprises 120,000 images categorized into six distinct classes. As mentioned in the previous section, this dataset is balanced, ensuring that each of the six categories contains an equal number of samples. In the initial version of the dataset introduced by the authors, 60,000 samples were allocated for training and 60,000 for testing. However, in our initial experiments, we modified this allocation by splitting the dataset into three parts: 60,000 samples for training, 30,000 for validation, and 30,000 for testing.

This approach allows us to optimize model performance through validation while ensuring robust evaluation on a separate test set. However, we believed that allocating such a large number of samples for the validation and test sets was unnecessary. Therefore, we modified the dataset split once more to create a new benchmark. In our benchmark, we allocated 90,000 images for training, 15,000 for validation, and 15,000 for testing. Additionally, we implemented appropriate functions to ensure that our modified dataset remains balanced across the different classes, which can be seen in Figure 6, facilitating effective training and evaluation of our jamming detection models.

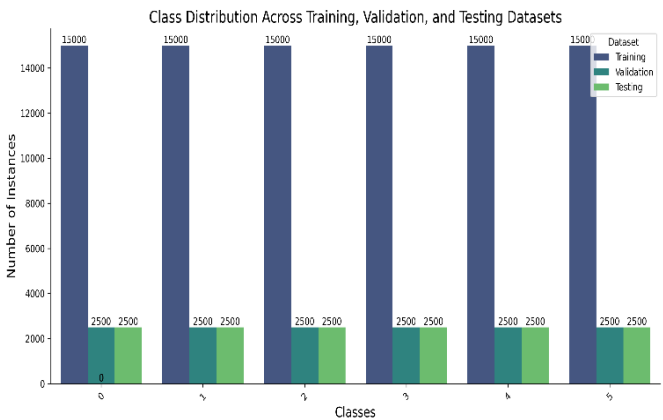

**Figure 6:** Balanced class distribution in the modified dataset, ensuring effective training and evaluation of our jamming detection models.

After splitting the dataset, we implemented data augmentation techniques, including resizing and normalizing, to enhance the quality and diversity of our training samples. These techniques not only improve the robustness of the models but also help prevent overfitting by providing a wider range of variations in the input data. In our initial experiments for jamming classification, we utilized several machine learning models, specifically Random Forest, Decision Tree, and Logistic Regression. These models were chosen for their effectiveness in classification tasks and their ability to handle different types of data distributions. To further improve our model performance, we implemented a feature extraction function using the ResNet18 model. This deep learning architecture is known for its ability to capture intricate patterns in image data through its residual learning framework.After extracting features from the images with ResNet18, we employed these features to train the aforementioned machine learning models. This approach allowed us to leverage the powerful representation capabilities of CNNs while still utilizing traditional machine learning algorithms for classification. By combining deep feature extraction with established machine learning techniques, we aimed to achieve a more accurate and efficient solution for jamming detection. The confusion matrix for random forest and logistic regression models can be seen in Figure 7.

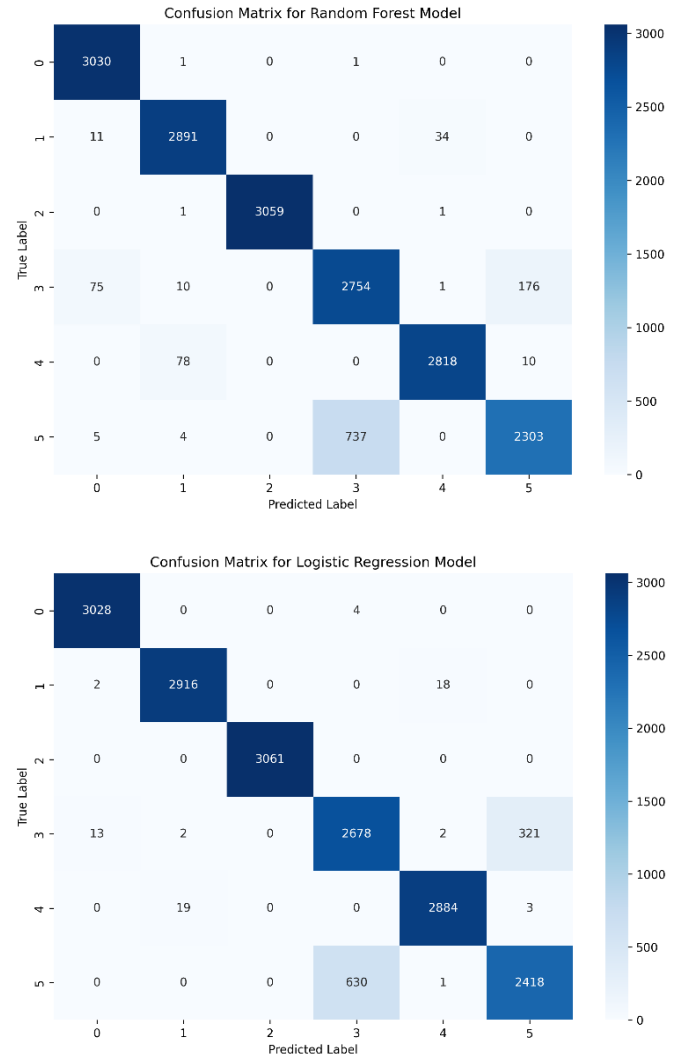

**Figure 7**: Confusion matrix for the Random Forest and Logistic Regression models, illustrating their classification performance.

With these simple modifications, we improved the average accuracy of the model by approximately 1.2% compared to the models reported in [Jammer classification in GNSS bands via machine learning]. This enhancement demonstrates the effectiveness of our approach in optimizing performance through data augmentation and feature extraction, leading to more reliable jamming detection outcomes.

However, for solving image classification tasks, this approach is not the optimal solution. Therefore, we modified our strategy by designing custom Convolutional Neural Network (CNN) models. CNNs are not only specifically tailored for processing image data, but they also excel at automatically learning spatial hierarchies of features through their convolutional layers. This capability allows CNNs to capture complex patterns and relationships within the images more effectively than traditional machine learning models. In our initial experiments, we designed a simple CNN model that incorporates residual blocks and contains approximately 6 million parameters. This architecture allows for improved training efficiency and better performance by enabling the network to learn more complex representations while mitigating issues such as vanishing gradients. We trained the model for 15 epochs using a one-cycle learning rate scheduler, which helps optimize the training process by adjusting the learning rate dynamically. As a result of this training regimen, we achieved an accuracy of 97.78%.

**Table 1:** Comparison of average accuracy (%) for various models employed in jamming detection, including traditional machine learning algorithms and deep learning architectures, highlighting the performance improvements achieved with ResNet18 using transfer learning.

| Model Name | Average Accuracy(%) |
|---|---|
| Best previous model based on the literature [32] | 94.90 |
| Logidtic Regression | 94.21 |
| Random Forest | 94.47 |
| XGBoost | 94.05 |
| Base Model | 97.78 |
| ResNet18(pretrained=False) | 98.56 |
| **ResNet18(pretrained=True)** | **98.93** |

To further improve the results, we trained the ResNet18 model both with and without transfer learning. This approach allows us to leverage pre-trained weights from the ImageNet dataset, which can enhance performance on our specific jamming detection task. By fine-tuning the model with transfer learning, we aimed to benefit from the rich feature representations learned from a large and diverse dataset, while also evaluating the effectiveness of training the model from scratch to understand the impact of transfer learning on our results. Table 1 presents a comprehensive comparison of average accuracy across various models employed for jamming detection, highlighting the effectiveness of both traditional machine learning and deep learning approaches. The best previous model reported in the literature achieved an accuracy of 94.90%, serving as a benchmark for subsequent evaluations. Among the traditional machine learning models, Logistic Regression, Random Forest, and XGBoost obtained accuracies of 94.21%, 94.47%, and 94.05%, respectively, indicating that while these models performed reasonably well, they were limited in their ability to capture the complexities of image data. In contrast, the Base Model, a simple CNN architecture, significantly improved performance with an accuracy of 97.78%. Notably, the ResNet18 model demonstrated exceptional results, achieving an accuracy of 98.56% when trained from scratch and 98.93% with transfer learning. These findings underscore the superior capability of deep learning models, particularly those utilizing transfer learning, to effectively classify jamming signals, thus emphasizing their potential for practical applications in this domain. Overall, the results highlight the importance of model selection and architecture in achieving high accuracy in jamming detection tasks.

For further information about our project, readers are encouraged to refer to the GitHub page where additional resources and documentation are available https://github.com/alicmu2024/GNSS-Jamming-Detection-and-Classification-using-Machine-Learning-Deep-Learning-and-Computer-Vision/tree/main.

## 6. CONCLUSIONS

Recently In conclusion, this research highlights the critical need for effective detection and mitigation strategies against jamming and spoofing threats to Global Navigation Satellite Systems (GNSS), particularly GPS. As reliance on GNSS technologies continues to grow across various sectors, the vulnerabilities associated with these systems become increasingly concerning. Our study employed advanced machine learning and deep learning techniques to address these challenges, achieving state-of-the-art results in both jamming and spoofing detection. Notably, we attained approximately 99% accuracy in jamming detection, marking a significant improvement of around 5% over previous studies. The extensive experiments conducted on real-world datasets underscore the potential of these technologies to enhance the integrity of GNSS systems. By making our code and models freely available on GitHub, we aim to foster further research and development in this vital area, contributing to the ongoing efforts to secure navigation systems against malicious threats. This work not only advances the field of GNSS security but also emphasizes the importance of integrating robust detection methodologies to safeguard critical infrastructure from emerging risks.